\documentclass[12pt,a4paper]{article}

\usepackage{amsmath,amssymb,amsfonts}
\usepackage[dvips]{graphicx}
\usepackage{epsfig}

\usepackage{color}

\usepackage{cite}
\usepackage{amsmath,amssymb,amsfonts,graphicx}
\usepackage{epsfig}
\usepackage[left=3.2cm,top=3.6cm,right=3.2cm,bottom=3.6cm,bindingoffset=0cm]{geometry}

\newcommand{\beq}{\begin{eqnarray}}
\newcommand{\eeq}{\end{eqnarray}}
\newcommand{\bea}{\begin{eqnarray}}
\newcommand{\eea}{\end{eqnarray}}

\newcommand{\bec}{\begin{center}}
\newcommand{\eec}{\end{center}}

\def\Tr{{\rm Tr}}

\numberwithin{equation}{section}

\begin{document}

\begin{flushright}\ \vskip -1.5cm {\small
    IFUP-TH\\NORDITA-2014-91}\end{flushright} 
\vskip 10pt
\begin{center}
{\bf\LARGE Vortex Zero Modes, Large Flux Limit and Ambj{\o}rn-Nielsen-Olesen  \\ 
\vskip 4pt
Magnetic  Instabilities}
\vskip 20pt
Stefano Bolognesi$^{(1,2)}$, Chandrasekhar Chatterjee$^{(2,1)}$, \\
 Sven Bjarke Gudnason$^{(3)}$ and  Kenichi Konishi$^{(1,2)}$\\[20pt]
{\em \normalsize
$^{(1)}$Department of Physics, E. Fermi, University of Pisa}\\[0pt]
{\em \normalsize
Largo Pontecorvo, 3, Ed. C, 56127 Pisa, Italy}\\[3pt]
{\em \normalsize
$^{(2)}$INFN, Sezione di Pisa,    
Largo Pontecorvo, 3, Ed. C, 56127 Pisa, Italy}\\[3pt]
{\em \normalsize
$^{(3)}$Nordita, KTH Royal Institute of Technology and Stockholm University,}\\[0pt]
{\em \normalsize  Roslagstullsbacken 23, SE-106 91 Stockholm, Sweden}\\[3pt]
{ \normalsize Emails: stefanobolo@gmail.com, chatterjee.chandrasekhar@pi.infn.it,}\\
{ \normalsize sbgu@kth.se, konishi@df.unipi.it}
\end{center}
\vskip 10pt
\date{July 2014}
\vskip 0pt

\begin{abstract}

In the large flux limit vortices become flux tubes with almost constant magnetic field in the interior region.
This  occurs in the case of non-Abelian vortices as well, and the study of such configurations allows us to reveal a close relationship between vortex zero modes and the gyromagnetic  instabilities of vector bosons in a strong background magnetic field  discovered by Nielsen, Olesen and Ambj{\o}rn.  The BPS vortices are exactly at the onset of this instability, and the dimension of their moduli space  is precisely reproduced in this way.  We present a unifying picture in which, through the study of the linear spectrum of scalars, fermions and $W$ bosons in the magnetic field  background,  the expected number of  translational, orientational, fermionic  as well as semilocal zero modes is correctly reproduced in all cases.

\end{abstract}
\newpage

\section{Introduction}

We  discuss some aspects of Abelian and non-Abelian vortex zero modes  in the large magnetic flux limit, and their relationship with the magnetic instabilities first studied in a series of papers by Nielsen, Olesen and Ambj{\o}rn \cite{Nielsen:1978rm,Ambjorn:1988tm,Ambjorn:1989bd}.

Our quest begins with the following observation.
The non-Abelian vortex is a generalization of the ordinary
Abrikosov-Nielsen-Olesen (ANO) vortex that carries non-Abelian
magnetic flux and supports  internal orientational zero modes
\cite{Auzzi:2003fs,Hanany:2003hp,Shifman:2004dr,Eto:2006pg}. Basically
it can be thought of as an ANO vortex embedded in a certain color-flavor corner, even though their moduli spaces and the dynamics of their fluctuations are found to be remarkably rich. 
On the other hand, it has been known for a long time that a non-Abelian magnetic field can trigger an instability in the presence of charged $W$-bosons that can become effectively tachyonic \cite{Nielsen:1978rm,Ambjorn:1988tm,Ambjorn:1989bd}. So the natural question is if these instabilities occur in the core of the non-Abelian vortex at all, and how they are related to the orientational zero modes of the latter.

It turns out that a natural setup to answer these questions is that of vortices in the large magnetic flux limit \cite{wallvortex1,wallvortex2,Bolognesi:2005rj,Bolognesi:2006pp}. In this limit, the profile functions simplify drastically, and the vortex becomes essentially a tube with constant magnetic field in the interior region separated from the vacuum by a thin domain wall. 
This solution resembles most the case of a constant magnetic field
background, which is the common situation considered in the early
works of the magnetic instabilities. We show that, for BPS vortices, no magnetic instability occurs. The magnetic field in the vortex interior is equal to the critical magnetic field and thus the effective mass of the lowest $W$-boson states is zero. This equivalence suggests that these states are related to the internal orientational zero modes.
The counting of the number of zero modes, discussed below, confirms this conjecture.

It will be shown that a generic interpretation holds for vortex zero modes in the large flux limit. They can be interpreted as charged fields (scalars, fermions or vector bosons) trapped inside the vortex in the lowest Landau level. 
The mechanism behind the generation of the zero modes is the cancellation between different contributions to the energy squared: 
the term from the lowest Landau level, 
the gyromagnetic term (this one is present only for vector bosons and fermions), 
and the bare mass squared. 
This analysis is applicable to all zero modes: translational,  orientational, fermionic and semi-local.

The paper is organized as follows. In Section \ref{abelian}, we review
the large flux limit of Abelian vortices and compute the translational
zero modes. We also show the existence of a domain wall separating the
two phases.  A related analysis of hole-vortex configurations nicely illustrates the relation between certain scalar zero modes in the linearized approximation  and the exact  translational 
zero modes of BPS  vortices.   In Section \ref{nonabelian}, we discuss the non-Abelian
vortex, its large flux limit, and analyze all types of  vortex zero modes,  gauge boson, scalar and fermion modes.  It is shown  
that, on the one hand, they arise with  exactly  the same mechanism as in the  onset of  general Ambj{\o}rn-Nielsen-Olesen instabilities, 
and  that, on the other,   their total  number  coincides in all cases studied, with the known dimension of the BPS  non-Abelian vortices or  with  the known index theorem. 
In Section  \ref{Discussion} we discuss  the significance of our results, and argue why  
the subtle relations found here between  two seemingly unrelated phenomena of  Ambj{\o}rn-Nielsen-Olesen instabilities and non-Abelian vortices, are nontrivial  and interesting. 
As an example of implications of our analysis,   we make a remark on some physics interpretation of Ambj{\o}rn and Olesen \cite{Ambjorn:1988tm,Ambjorn:1989bd}.

\section{The Abelian vortex}
\label{abelian}

We first review the large flux limit of Abelian vortices \cite{wallvortex1,Bolognesi:2005rj}.
We consider the Abelian-Higgs model
\begin{equation}
\label{abelianvortex}
\mathcal{L}=-\frac{1}{4}F_{\mu\nu}F^{\mu\nu} + |(\partial_{\mu}-i e A_{\mu}%
)q|^{2}-V(|q|)\ ,
\end{equation}
with the following potential 
\beq
V= \frac{ \lambda^2 e^2}{2} (|q|^2 -{\xi})^2\ ,
\eeq
whose minimum 
$|q|=\sqrt{\xi}\neq0$ is in the Higgs phase.
The choice of $\lambda =1$ corresponds to  having a BPS potential.

The fields for an axially symmetric vortex of charge $n$ can be
parametrized by the following Ansatz
\begin{align}
\label{vortex}
q  &  = \sqrt{\xi} e^{in\theta}\,q(r)\ ,\\
A_{\theta}  &  =\frac{n}{e r}\,A(r)\ .\nonumber
\end{align}
The  profile functions $q(r)$ and $A(r)$
 are subject to the boundary conditions $q(0)=0$, $q(\infty)=\sqrt{\xi}$ and $A(0)=0$,
$A(\infty)=1$.
 The claim of \cite{wallvortex1,wallvortex2} is that, for every Higgs-like potential $V$, in the
large-$n$ limit the profile for the scalar field converges to a step function
\begin{align}
\label{limq}
\lim_{n\rightarrow\infty}q(r) = \theta_{H}(r-R_{bag})  \ ,
\end{align}
where $\theta_H$ is the Heaviside step function and the vortex radius,
$R_{bag}$, will be determined shortly.   
The gauge field profile 
converges to the following limit
\beq
\label{gaugeinside}
\lim_{n\rightarrow\infty}A(r) = \left\{\begin{array}{cc} 
\frac{r^2}{ R_{bag}^2} & r \leq R_{bag} \ , \\
1 & r > R_{bag} \ .
\end{array}\right. 
\eeq
The magnetic field is zero outside the bag and constant inside
\beq
B|_{r\leq R_{bag}} = \frac{2 n}{e R_{bag}^2} \ , \qquad\qquad 
B|_{r > R_{bag}} =0 \ .
\eeq
The  total  magnetic flux is fixed by the boundary condition
\beq
\Phi_B = \oint A_{\theta} = \frac{2 \pi n}{e} \ .
\eeq
This conjecture has been shown to hold numerically with great
precision in \cite{Bolognesi:2005rj}. 
The step function of the profile $q(r)$ reveals the presence of a
substructure: a domain wall interpolating between the Coulomb phase $q=0$ and
the Higgs phase $|q|= \sqrt{\xi}$. This wall has a physical thickness
which is an ${\cal O}(1/\sqrt{n})$ effect respect to the bag radius.

The radius of the bag is determined by minimization of the tension.
The tension has two contributions, one from the magnetic field and one
from the potential energy at $q=0$, i.e.~inside the bag
\beq
T(R) = \frac{2 \pi n^2 }{ e^2 R^2} + \frac{\lambda^2 e^2 \xi^2 \pi
  R^2}{2} \ , 
\eeq
and its minimization gives
\beq
\label{rbag}
R_{bag}^2 = \frac{2 n}{ \lambda e^2 \xi} \ .
\eeq
The tension of the vortex is then
\beq
T_{bag} = 2 \pi n \lambda \xi \ .
\eeq
The value of the magnetic field $B$ inside the bag is  
\beq
\label{Binside}
   B = \lambda e \xi \ .
\eeq
Note that $B$ is independent of $n$.

We now want to study the spectrum of fluctuations around this solution.
The phase outside the vortex is gapped, with the photon mass $e\sqrt{2\xi}$ and the scalar mass $\lambda e\sqrt{2\xi}$. 
The phase inside is more interesting. Here we have a massless gauge field
with a background constant magnetic field (\ref{Binside}) which is  coupled to a charged scalar field $q$. To compute the mass of the field we have to expand around the tip of the potential:  
\beq
V =  \frac{ \lambda^2 e^2 \xi^2}{2}  - \lambda^2 e^2 \xi |q|^2 + \frac{\lambda^2 e^2}{2} |q|^4 \ .
\eeq
This is a 'tachyon' with negative mass squared
\beq
m^2 = - \lambda^2 e^2 \xi \ . 
\eeq
The quartic term can be neglected in the limit of small fluctuations
\beq
\label{smallfluctuations}
\delta q \ll \sqrt{\xi} \ .
\eeq 
Tachyons are in general a signal of instabilities, but here we also
have to take into account the effect of the background magnetic field
before jumping to conclusions.

Inside the bag we choose the symmetric gauge for the gauge field,
viz.~$A_k = (-B y/2, +B x/2) $.
The scalar field equation,  in the limit of small fluctuations (\ref{smallfluctuations})),  is the linear equation
\beq
\big(\partial_t^2 - (\partial_x -ieA_x)^2 - (\partial_y -ie A_y)^2  + m^2\big) q = 0 \ .
\label{scalarenergy}
\eeq
Substituting
\beq
q = e^{i E t} \phi(x,y) \ ,
\eeq
 the energy-squared operator is then given by
\beq
\label{spectrumscalar0}
E^2 \phi = \big(- (\partial_x -i e A_x)^2 - (\partial_y -i e A_y)^2  + m^2 \big) \phi \ .
\eeq
The operator on the right-hand-side is the same as that of the
non-relativistic Landau level problem, and the same technique can be
used for its diagonalization. Changing to complex coordinates: 
$z=\sqrt{e B}\, (x+iy)$ and $\bar{z}=\sqrt{e B}\, (x-iy)$, the
spectrum operator can be rewritten as 
\beq
E^2 \phi  = \big( e B (a^\dagger a + 1) + m^2 \big) \phi \ ,
\eeq
with the operators $ a = z/2 + 2\partial_{\bar{z}}$ and  $ a^\dagger = \bar{z}/2 -2 \partial_z$ satisfying the  commutation relation $[a,a^\dagger]=2$. 
The eigenstates are then
\beq
\label{solstates}
\phi_{n_1,n_2} = {a^\dagger}^{n_2} (z^{n_1} e^{-|z|^2 /4}) \ .
\eeq
The  energy spectrum is then
\beq
\label{spectrumscalar1}
{E}_{n_1,n_2}^2 = (2n_2 +1) e B  + m^2 \ .
\eeq

The ground state, which is the lowest Landau level, has the energy
$E_0 = \sqrt{ e B + m^2}$.  If the scalar field is allowed to have a
tachyonic mass, then the ground state becomes massless at the critical
value $m^2 = -e B$. Below this point, two zeros of (\ref{spectrumscalar1})
disappear in the complex plane and the field becomes really tachyonic.  
This situation is precisely realized for vortices with large flux.
  Using (\ref{Binside}) and (\ref{spectrumscalar1}), the energy of the ground state is
\beq
\label{spectrumscalar2}
E_0 = e  \sqrt{\xi \lambda (1  -  \lambda)  } \ .
\eeq
Thus the spectrum is gapped for $\lambda < 1$, massless for $\lambda
=1$ and tachyonic for $\lambda >1$. This result has a nice physical
interpretation. Type I vortices, $\lambda < 1$, are known to attract
each other and this is manifested, in the large $n$ limit, by the
stability of the spectrum. For the type II vortices, $\lambda > 1$,
there is repulsion between the vortices and this is manifested in the
tachyonic instability of the multi-vortex.  We may then want to
interpret the massless state for $\lambda =1$ as the zero modes of BPS
vortices.

We check that  the number of zero modes is correctly reproduced. For
BPS vortex we have $2n$ zero modes corresponding to translations in
the transverse plane. The ground state Landau level, $n_2=0$, is not
isolated, but come with a degeneracy proportional to the area.   
The states (\ref{solstates}) are concentric rings localized at radius $R_{n_1} \simeq  \sqrt{2 n_1 /e B} $ so the density of zero modes per unit of area is $e B / 2 \pi$.
The number of zero modes in the area spanned by the bag is  then 
\beq
\label{scalarzeromodes}
\#_{zero \ modes} = R_{bag}^2 \frac{ e B }{2}  =  n \ .
\eeq
It is thus natural to associate them with the $n$  translational zero
modes of the BPS equations.   Note that for a BPS vortex of winding
number $n$, the dimension of its moduli space can be found
conveniently by going to the limit of far-distant $n$ minimal
vortices, whose translational moduli are simply given by ${\bf C}^{n}$.     
This approach has basically neglected the back reaction of the zero
modes on the gauge fields and on themselves via the quartic
interaction. The approximation is thus valid in the linear
approximation  of small fluctuations (\ref{smallfluctuations}).

In the large-$n$ limit the radius of the vortex (\ref{rbag})
goes to infinity while the magnetic field in the interior region (\ref{Binside}) remains fixed.
This suggests that the domain wall separating the two phases should
exists also in isolation, and as a proper wall it should be
translational invariant in one direction. We will now show that indeed
this object exists in isolation for the BPS theory.

\begin{figure}[h!t]
\centerline{
\epsfxsize=6.0cm \epsfbox{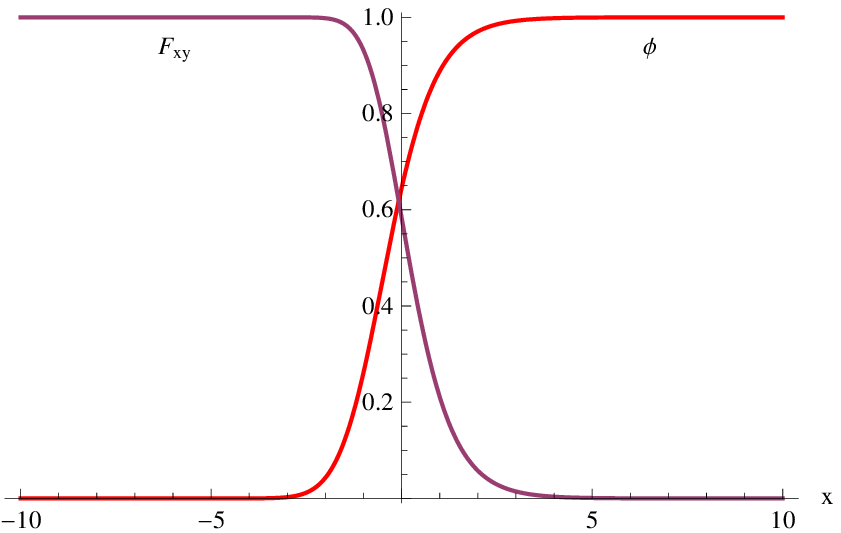} \qquad \quad 
\epsfxsize=6.0cm \epsfbox{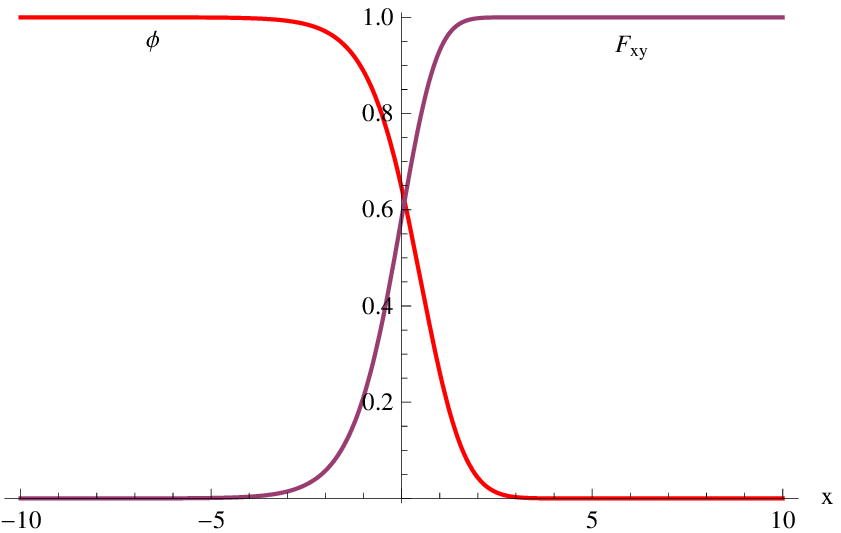}}
\centerline{
\epsfxsize=6.0cm \epsfbox{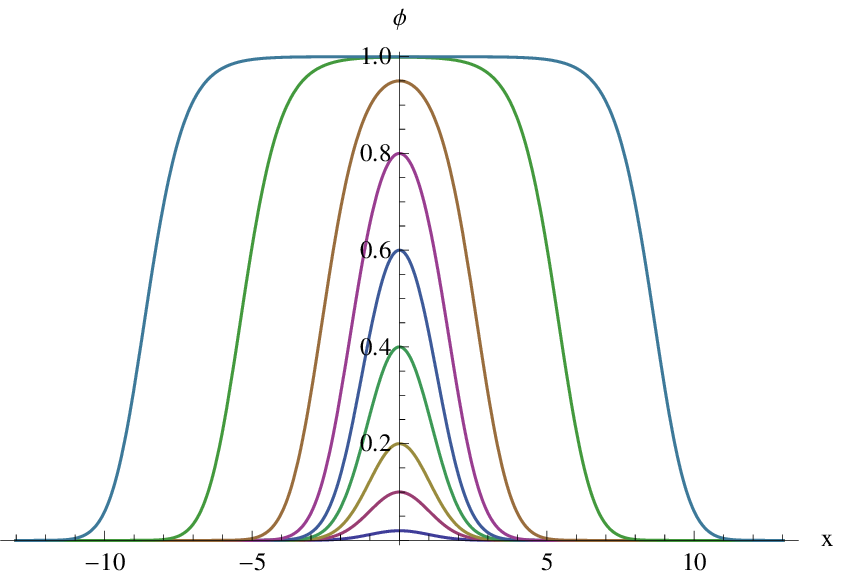} \qquad \quad
\epsfxsize=6.0cm \epsfbox{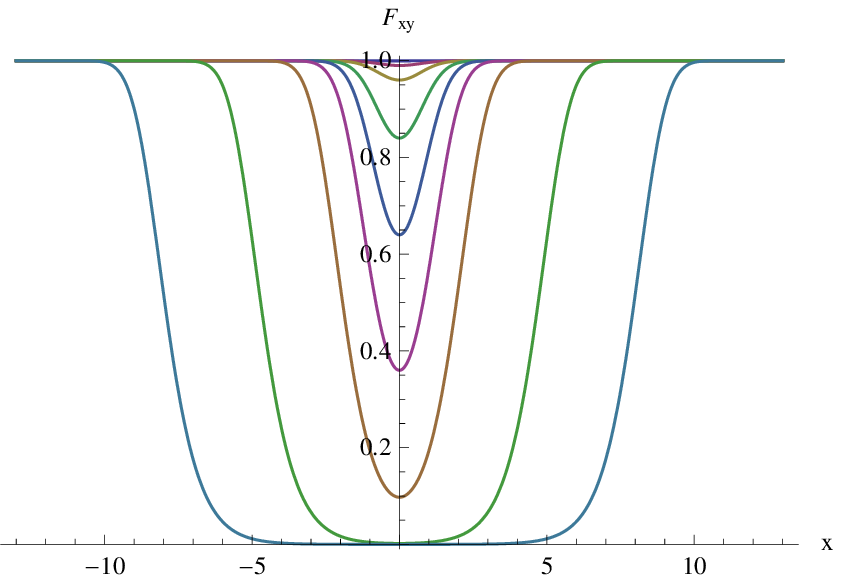}}
\caption{{\footnotesize Top row: The two domain wall solutions of 
    Eq.~(\ref{eqphiwall}) with $e=\xi=1$. Bottom row: A
    one-parameter family of solutions with two walls at various 
    distances: (left) the solutions $\phi$ and (right) the
    corresponding magnetic fields $F_{xy}$.}}  
\label{walls}
\end{figure}

The  Bogomol'nyi completion of the static energy density is
\bea
\mathcal{H} &=& \frac{1}{2}\left[ F_{xy} + e(|\phi|^2-\xi) \right]^2
+\left|(D_x + i D_y)\phi\right|^2  \nonumber \\
&& + \, e \, \xi F_{xy} 
- i\varepsilon^{ij}\partial_i\left[(D_j \phi)\phi^\dag\right] \ . 
\eea
We take the domain wall to be extended in the $y$ direction. 
Furthermore, we choose to work in the analogue of the vortex singular gauge (the singularity for the wall is pushed to $y \to \pm \infty$), thus the scalar field is a function of $x$ only with no winding and we can set $A_x=0$. 
Writing down the BPS equations for $\phi(x)$ and $A_y(x)$ we have
\bea
A_y' + e(\phi^2 - \xi) &=& 0 \ , \nonumber \\
 \phi' + e  A_y \phi &=& 0 \ .
\eea
Solving for $A_y$ and plugging the second BPS equation into the first, we get
\beq
\label{eqphiwall}
(\log\phi)'' = e^2(\phi^2 - \xi) \ .
\eeq
In the first row of Figure \ref{walls} are shown two numerical
solutions to this equation.  
There are two domain walls separating the Higgs phase from a Coulomb
phase with constant magnetic field. Note that the two walls are both
solutions to the same BPS equation, they are related by parity and
charge conjugation. 
Solutions with arbitrary separation between  the two walls are also
possible and are displayed in the second row of Figure \ref{walls}.

\begin{figure}[h!t]
\centerline{
\epsfxsize=12.0cm \epsfbox{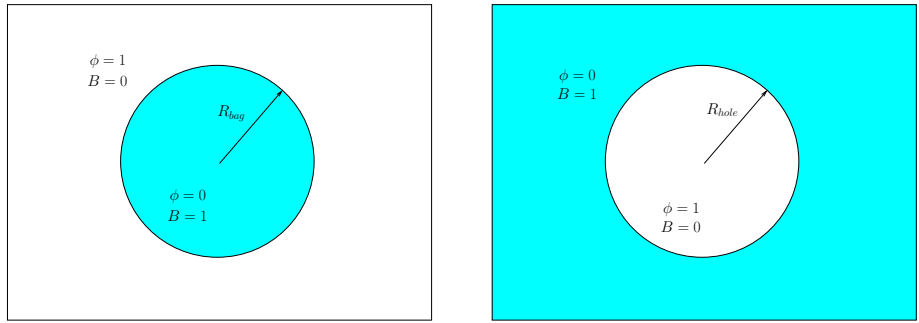}}
\caption{{\footnotesize Vortex bag (left) compared with the
    hole-vortex (right).}}
\label{holevortexfig}
\end{figure}

\begin{figure}[ht!]
\centerline{
\epsfxsize=6.0cm \epsfbox{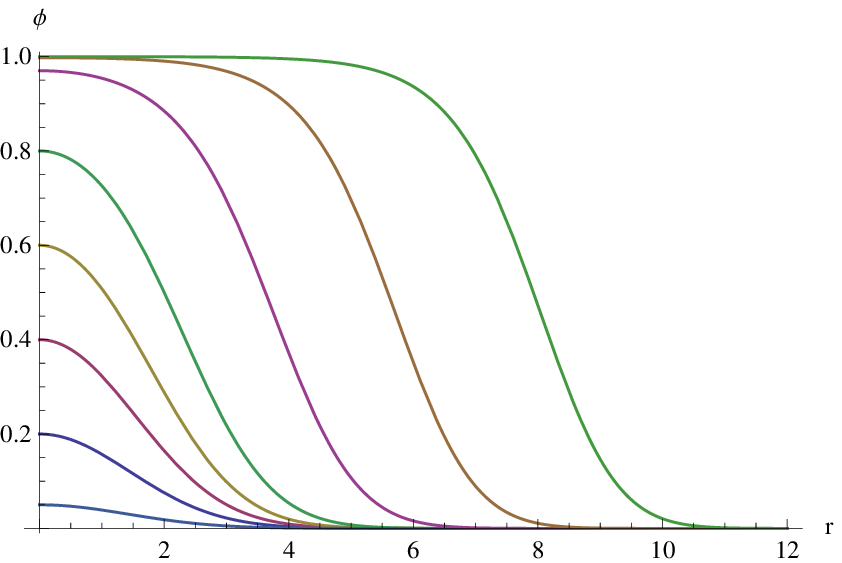}\qquad \quad
\epsfxsize=6.0cm \epsfbox{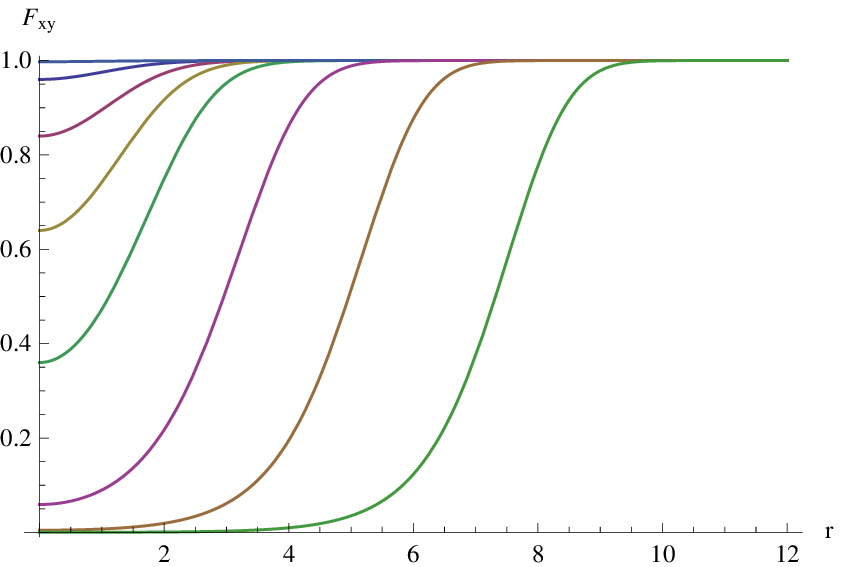}}
\caption{{\footnotesize Examples of a one-parameter family of
    solutions for the hole-vortex of
    Eq.~(\ref{eqphiradialholevortex}): (left) the solutions $\phi$ and 
    (right) the corresponding magnetic fields $F_{xy}$. }}
\label{radial}
\end{figure}

Let us consider a final configuration, which clarifies the relation
between the domain wall solutions of Figure \ref{walls} and the linear
zero modes previously discussed. 
We consider a `hole-vortex', which is a region of zero magnetic field
in a background of constant magnetic field (see Figure
\ref{holevortexfig}). The  axial-symmetric Ansatz  is  
\bea
 A_{\theta} = \frac{e \xi r}{2} - \frac{1}{r}f(r) \ ,  \qquad  
 A_r = 0 \ , \qquad 
 \phi =  \phi(r) \ , 
\eea  
with boundary conditions $\phi(\infty) = 0$ and  $\phi'(0)=0$ for the Higgs field.
The value of $\phi(0)$ is left to be arbitrary. Note one difference between the vortex and the hole-vortex. The missing flux inside the hole vortex, which is $2\pi \int_0^{\infty} dr f'$ and is related to $\phi(0)$,  is a continuous parameter: it is not quantized. Inserting the Ansatz into the BPS equations we obtain
\bea
-\frac{ f'}{r} + e\phi^2  &=& 0  \ , \nonumber  \\
\phi' + e \left(\frac{e\xi r}{2} - \frac{f}{r}\right) \phi &=& 0  \ .
\eea
From this we obtain a second-order equation for $\phi$:
\beq
\label{eqphiradialholevortex}
 \frac{1}{r} (r (\log\phi)')'  =  e^2(\phi^2 - \xi) \ .
\eeq
Both from analytic inspection of the equation, and from the shape of
the numerical solutions, we can detect two different regimes. When
$\phi(0)$ is very small, the $\phi^2$ term on the right-hand side of
Eq.~\ref{eqphiradialholevortex} is negligible and the solution is thus  
\beq
\phi \simeq e^{-e^2 \xi r^2/4} \;; \qquad \quad 
F_{xy} = e\xi - {\cal O}(\phi^2) \ :   \label{tiny} 
\eeq
this is exactly the first Landau level (\ref{solstates}) with $n_1 = n_2 =0$.
The magnetic field does not receive any correction to linear order in $\phi$.
When $\phi(0) \simeq \sqrt{\xi}$ the hole-vortex is well approximated by a ring of domain wall as equation (\ref{eqphiradialholevortex}) becomes almost equivalent to (\ref{eqphiwall}).  Examples are shown in Figure \ref{radial} \footnote{In contrast to the vortex (the left of Fig. 2),  the hole-vortex (the right figure) does not represent a minimum-tension configuration, as it stands. 
For its stability, it is necessary to consider the external  region with  $B \ne 0$  as a part of a vortex  with a  fixed quantized total flux.  This makes  perfect sense, as the tiny hole-vortex 
(\ref{tiny}) can then be thought of as a germ of the instability of the vortex itself,  occurring anywhere inside the vortex }.

\section{The non-Abelian vortex}
\label{nonabelian}

For a generic particle with spin $S$ and gyromagnetic ratio  $g_S$ the spectrum in a constant magnetic field is:
\beq
\label{spectrumgeneric}
{\cal E}_{n_1,\vec{S}}^2 = (2n_1 +1) e  B  +  g_S e  \vec{B} \cdot \vec{S} + m^2 \  ,
\eeq
where $g_S e  \vec{B} \cdot \vec{S}$ is the Zeeman term.
For Dirac fermions we have $S=1/2$ and $g_S =2$ and the spectrum is 
\beq
\label{spectrumfermion}
{\cal E}_{n_1, \uparrow \downarrow}^2 = (2n_1 +1) e B  \pm   e  B  + m_{F}^2 \ .
\eeq
The Zeeman term, for the right choice of spin orientation, cancels exactly the first Landau level term.  This is the reason for the existence of fermionic zero modes, whenever $m_{F}=0$.
The generalization for charged spin-1 $W$ bosons will be of interest in  the rest of this section.

We now consider the theory of non-Abelian vortices \cite{Auzzi:2003fs}.
Stripped to its basic constituents, the model consists of a  $U(N)$ gauge theory coupled to $N$ flavors of fundamental quarks
\beq
\label{nonabelianvortex}
{\cal L} = -\frac{1}{2} \Tr_{N}  (F_{\mu\nu}F^{\mu\nu}) + \Tr_N (D_\mu
q)(D^\mu q)^{\dagger} - \frac{\lambda^2 g^2}{4} \Tr_{N} \left(qq^\dag -{\xi}{\bf 1}_{N \times N}\right)^2\ .
\eeq
with $D_{\mu} = \partial_{\mu} -i g A_{\mu}$. For the moment we
consider the case of equal couplings for the $U(1)$- and $SU(N)$-part
of the gauge group. 
The vacuum is the color-flavor locked phase
\beq
q=\left(\begin{array}{ccc}
\sqrt{{\xi}}&&\\
&\ddots&\\
&&\sqrt{{\xi}}\\
\end{array}\right) \ .
\label{qvacuum}
\eeq
The color-flavor diagonal $U(N)$ symmetry is unbroken by the vacuum.
The mass of the gauge bosons in the vacuum is $M^2 =  g^2 \xi$, and this is true for all the generators of the $U(N)$ gauge group.

The $SU(N)\times U(1)$ gauge symmetry is completely broken and, as $\pi_1(SU(N)\times U(1)) = {\bf Z}$, the system supports vortices. 
To build a vortex configuration we embed  the ordinary Abelian  $U(1)$ vortex in this theory. A minimum individual vortex configuration breaks the residual symmetry to  
$SU(N-1)\times U(1) \subset SU(N)$ and the vortex acquires orientational zeromodes of the coset $\mathbf{C}P^{N-1}$.   A  possible Ansatz for a multi-vortex of charge $n$ is 
\bea
\label{vortexconf}
q&=&\left(\begin{array}{cccc}
e^{in\theta}\sqrt{\xi}q(r)&&&\\
&\sqrt{\xi}&&\\
&&\ddots&\\
&&&\sqrt{\xi}\\
\end{array}\right)\ ,   \nonumber  \\
A_k&=&\left(\begin{array}{cccc}
-\epsilon_{kl} n \hat{r}_l A(r)/g r&&&\\
&0&&\\
&&\ddots&\\
&&&0\\
\end{array}\right) \ . \label{particular}
\eea
This corresponds to having $n$ non-Abelian vortices in the same spatial position and in the same internal orientation.  It is only a special point in the big moduli space of $n$ non-Abelian vortices, but for the moment is the one we shall focus on.
Since this is just an embedding of the ANO axial-symmetric vortex
(\ref{vortex}), the same considerations about the large-$n$ limit
discussed in the previous section hold (this is valid only in the case
of equal couplings for $U(1)$ and $SU(N)$). 
In particular the large-$n$ limit of the profile functions is  (\ref{limq}) and (\ref{gaugeinside}).  
 So we may use all the formulae of the previous section by replacing
 $A_{\mu} \to \sqrt{2} A_{\mu}$ and $e$ with $g/\sqrt{2}$ to account
 for the different normalization of the generators.

\begin{figure}[!th]
\centerline{
\epsfxsize=12.0cm \epsfbox{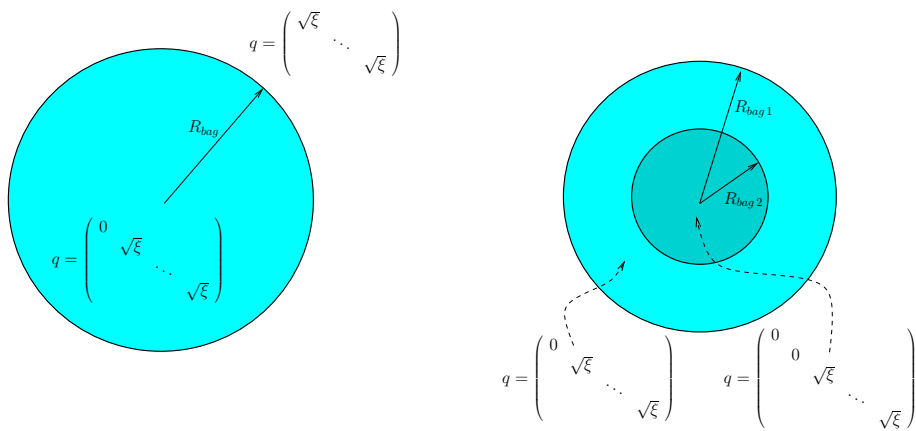}}
\caption{{\footnotesize Two possible configurations of large-$n$
    multi-vortices.}} 
\label{vortexfig}
\end{figure}

We are  interested in the spectrum around this multi-vortex which is
sketched in the left of Figure \ref{vortexfig}. Outside the bag radius
the scalar fields take the form of Eq.~(\ref{qvacuum}) and all the
states, gluons and scalars, are massive. 
Inside the bag, however, the scalar quarks are
\beq
\label{qinsidena}
q=\left(\begin{array}{cccc}
0&&&\\
&\sqrt{\xi}&&\\
&&\ddots&\\
&&&\sqrt{\xi}\\
\end{array}\right)\ .
\eeq
The radius of the bag is given by
\beq
R_{bag}^2 = \frac{4 n}{ \lambda g^2 \xi} \ ,
\eeq
and the value of the $B$ field is constant inside the bag and given by 
\beq
\label{binsidena}
   F_{xy} =\left(\begin{array}{cccc}
\lambda g \xi /2&&&\\
& 0 &&\\
&&\ddots&\\
&&&0\\
\end{array}\right)\ .
\eeq
The field $q_{11}$ has a negative mass squared, and its spectrum is the same as in the Abelian case. In particular, for $\lambda =1$ this field gives $n$ complex zero modes to be associated with the translational zero modes of the vortex. All the fields in the reduced sector $(N-1)^2$ are massive, as they are in the vacuum state. 

The interesting thing happens for the $N-1$, $W$ bosons in the
following matrix components of the gauge fields 
\beq\left(\begin{array}{cccc}
&*&\dots&*\\
*&&&\\
\vdots&&&\\
*&&&\\
\end{array}\right)\ .  \label{corners}
\eeq
These are charged particles and thus they couple to the magnetic field inside the vortex. We are interested in computing the spectrum for those.  We can consider the problem of $N=2$ where we have to deal with one $W$ boson only. 
So we denote
\beq
A_{\mu} =\left(\begin{array}{cc} A_{\mu} & W_{\mu} \\ W^*_{\mu} & B_{\mu} \end{array}\right) \ ,  \qquad \qquad  q=\left(\begin{array}{cc}q_{11} & q_{12} \\  q_{21} & q_{22} \end{array}\right) \ .
\eeq
The terms in the Lagrangian which contributes to the $W$ mass are
\beq
\label{massw}
g^2 |W_{\mu}|^2 (|q_{11}|^2 + |q_{22}|^2) = \left\{ \begin{array}{cc}
  g^2 \xi |W_{\mu}|^2  & r \leq R_{bag} \ ,\\ & \\ 2  g^2 \xi |W_{\mu}|^2
  & r > R_{bag} \ . \end{array} \right. 
\eeq
The $W$ boson is massive everywhere, but the mass squared inside the bag is reduced by half since only $q_{22}$ contributes to the mass term:  this fact will be very important below.
The Lagrangian, reduced to the $W$-boson sector, is 
\bea
{\cal L} &=& -\frac{1}{2} F_{\mu \nu} F^{\mu\nu} -  |D_{\mu} W_{\nu} - D_{\nu}W_{\mu}|^2   \nonumber \\ & -&    2 \,i\, g\,  F_{\mu\nu}    W^{\mu *} W^{\nu} +2 \, m_W^2 \, |W_{\mu}|^2+ {\cal O}(g^2 W^4) \ , 
\eea
where
\beq  
\label{massWbosoninside}
 m_{W}^{2}=  \frac{g^2 \xi }{2} \ .
\eeq
The quartic term can be neglected for small fluctuations 
\beq
\delta W \ll \sqrt{\xi} \ .
\eeq
The linear equation for the $W$ boson, in the gauge $D_{\mu} W^{\mu} =0$, is 
\beq
\label{eqwboson}
\Big( (D^\rho D_{\rho} + m_W^2) \eta_{\mu\nu} - 2 i g F_{\mu\nu} \Big) W^{\nu} = 0  \ .
\eeq
It is important to note that the $W$ boson is {\it not} minimally coupled to the gauge field $A_{\mu}$.
For minimally coupled fields the gyromagnetic factor is $g_S = 1/S$ while for the $W$ boson, $g_S =2$ and not $1$.  This is due to the last term in (\ref{eqwboson}).  
We consider the magnetic field directed in the third direction $F_{12} = B$.

The solution we are mainly interested in is given by the following
\beq
W_{\mu} = e^{i E t} \begin{pmatrix} 0 \\ w(x,y) \\  i w(x,y)
  \\ 0\end{pmatrix} \ ,\label{profile}
\eeq
which is the negative eigenstate of the spin in the magnetic field
direction
\beq
(S_3)_{\mu\nu} W_{\nu} = - W_{\mu} \, . \nonumber
\eeq
For these states the spectrum is
\beq
E^2 w =  \big(- (\partial_x -i g A_x)^2 - (\partial_y -i g A_y)^2  - 2
g  B  +  m_W^2 \big) w \ ,
\eeq
and the gauge fixing condition becomes 
\beq
-D_{\mu} W^{\mu} = e^{i E t}\left(\partial_x +i \partial_y + \frac{g B}{2} (x + iy)\right) w = 0 \ .  \label{zeroenergy}
\eeq
The solution is then given by the lowest Landau level states
\beq
\label{solstatesna}
w = f(z) e^{-|z|^2 /4} \ , 
\eeq
with $f(z)$ any holomorphic function, and the spectrum for those states is
\beq
\label{spectrumW}
E^2 = -  g  B  + m_{W}^2 \ .
\eeq
For a generic state, with eigenvalue of $S_3$ which can be $\epsilon = \pm 1 , 0$ and Landau level $n_1$, the spectrum is 
\beq
{\cal E}_{n_1, \epsilon}^2 = (2n_1 +1) g B  + 2 \epsilon  g  B  + m_{W}^2 \ .   \label{perfect}
\eeq
Note that the non-minimal coupling of the $W$ boson is responsible for the anomalous gyromagnetic factor $g_s =2$.
Now the Zeeman term is twice the first Landau level term, and so the ground state energy is $ E_{n,-1} = \sqrt{ - g  B + m_{W}^2} $. 
This is somehow similar to the scalar field story of Section \ref{abelian}, except for the fact that the critical value for the existence of zero modes is now a positive mass squared, $m_{W}^2 = B$, and not a negative one. 
For  $m_{W}^2 < B$ we have an instability; the ground state becoming
tachyonic is the signal of a phase transition which can be driven by
the $W$ condensate.  
For pure Yang-Mills (i.e. $m_W =0$) the ground state is always tachyonic for any $B \neq 0$. This  is the instability discussed by Nielsen, Olesen and Ambj{\o}rn  \cite{Nielsen:1978rm,Ambjorn:1988tm,Ambjorn:1989bd}.

The ground state energy for the $W$ bosons, taking into account the value of the magnetic field (\ref{binsidena}) and the mass inside the bag (\ref{massw}),  is given by
\beq
\label{Wbarrier}
E_0 = g  \sqrt{\frac{\xi}{2}}  \sqrt{1- \lambda  } \ .
\eeq 
This  is sub-critical for $\lambda <1$ (type I vortices) and above-critical for $\lambda > 1$ (type II vortices).  This nicely fits with the expectation that for type I the ground state is given by  vortices all in the same orientation state while for type II this is an unstable point.
 For the BPS case $\lambda =1$ we have exactly $B = B_{cr}$.
The number of zero modes, including the scalars (\ref{scalarzeromodes}), $n$,   and the $W$ bosons  ($(N-1) n$),  in total is
\beq
\#_{zero \ modes} = N R_{bag}^2 \frac{g B }{2}  = N n \ ,   \label{determination}
\eeq
which is in agreement with the dimension of the moduli space of the winding number $n$, BPS non-Abelian vortices (i.e.~the number of the zero modes). Indeed, even though the  structure of the moduli space of higher-winding non-Abelian vortices 
is  quite rich  and has been studied only for some simplest cases \cite{Hashimoto:2005hi,Auzzi:2005gr,Eto:2006cx,Eto:2009bg,Gudnason:2010rm},  its dimension is known from the index theorem \cite{Hanany:2003hp,Eto:2009bg}.   Alternatively it can be deduced from the limiting case where the $n$ minimal vortices are well separated. The moduli space approaches 
in that limit  the form \cite{Eto:2005yh}
\beq      \left( {\bf C} \times \mathbf{CP}^{N-1} \right)^n/S_{n} \ ,   \label{than}
\eeq
where $S_{n}$ is a permutation  of $n$ vortices.  Its dimension is given by
\beq  
n (N-1 +1) = N n \ . \label{orientation}
\eeq
The determination of the number of zero modes (\ref{determination})
was made by studying the properties of the fluctuation of the
particular vortex solution (\ref{particular}). Around that point the
structure of the vortex moduli space is certainly more complicated
than (\ref{than}), but since the dimension of a manifold is the same
at any point,  the agreement between (\ref{determination}) and
(\ref{orientation})  shows that the zero modes related to the
orientational and translational zero modes of the BPS non-Abelian
vortices have indeed the same origin as the zero (or negative) modes
which trigger the Ambj{\o}rn-Nielsen-Olesen instabilities.

In a supersymmetric extension of our model, the fermions get mass through the Yukawa term, 
\beq     {\cal L}_{Yukawa} =    \sqrt{2}  g   \,  {\bar q}_{A}  \,
\lambda  \,\psi^{A}  + {\rm h.c.}\:,
\eeq
where $\lambda$ are gauge fermions in the adjoint representation of
the color gauge group, $SU(N)\times U(1)$ and $A=1,2,\ldots, N_{f}=N$ is the flavor index.   The scalar VEV
(\ref{qinsidena}) inside the vortex implies that the nonvanishing
Dirac mass terms are 
\beq   \sqrt{2}  g   \,  \sqrt{\xi}  \, \sum_{A=2}^{N}  \sum_{i=1}^{N}
(\lambda)_{A}^{i} \psi^{A}_{i} + {\rm h.c.}\:;
\eeq
note that  the fermions $ (\lambda)_{1}^{1}$ and   $\psi^{1}_{i}$,
$i=1,2, \ldots, N$ do not appear; they can be thought of as  
$N$  massless  Dirac fermions.  We see from  Eq.~(\ref{spectrumfermion})  that the number of the fermionic zero modes is then  $N  n
$,   as expected.

As a further nontrivial check, we
consider another multi-vortex configuration, i.e.~the one sketched on
the right of Figure \ref{vortexfig}. It consists of two multi-vortices,
one with radius $R_{bag\, 1}$ and the other with radius $R_{bag\, 2}$,
in mutually orthogonal internal orientations. Hence, in the theory
(\ref{nonabelianvortex}) which has equal couplings for the $U(1)$ and
the $SU(N)$ parts, they can overlap with no modification of their
profile functions. 
We take the two vortices to have respectively $n_1$ and $n_2$ units of flux, so we expect to recover a total of $N (n_1 + n_2)$ complex zero modes.
We take the second vortex to be completely immersed in the other one, as in Figure \ref{vortexfig},  with $n_1 > n_2$ and 
\beq
R_{bag\, 1}^2  = \frac{2 n_1}{ g^2 \xi} \ , \qquad \qquad 
R_{bag\, 2}^2  = \frac{2 n_2}{ g^2 \xi} \ .
\eeq
We now consider only the case $\lambda =1$.
In the ring between $R_{bag\, 2}$ and $R_{bag\, 1}$ the scalar field and magnetic field are the same of the previous example, (\ref{qinsidena}) and (\ref{binsidena}),  and the counting of zero modes is unchanged. We have one mode from the scalar field 
\beq
\#_{zero \ modes \ ring} = N \left( R_{bag \, 1}^2 -R_{bag \, 2}^2
\right) \frac{  g B }{2}  = N (n_1-n_2) \ . 
\eeq
In the internal disk we have instead the following fields
\beq 
\label{qinsidena2}
q=\left(\begin{array}{ccccc}
0&&&&\\
&0 &&&\\
&&\sqrt{\xi}&&\\
&&&\ddots&\\
&&&&\sqrt{\xi}
\end{array}\right) \ ,\qquad   F_{xy} =\left(\begin{array}{ccccc}
 g \xi /2&&&&\\
&g \xi /2 &&& \\
& &0 &&\\
&&&\ddots&\\
&&&&0\\
\end{array}\right)\ .
\eeq
The zero modes are $4$ scalars and $2 (N-2)$ $W$ bosons in the following components
\beq
\delta q = 
 \left(\begin{array}{ccccc}
*&*&&&\\
*&*&&&\\
&&&&\\
&&&&\\
&&&&\\
\end{array}\right) \ ,
\qquad \quad 
W = \left(\begin{array}{ccccc}
&&*&\dots&*\\
&&*&\dots&*\\
*&*&&&\\
\vdots&\vdots&&&\\
*&*&&&\\
\end{array}\right)\ ,
\eeq
so a total of $2N$. 
The number of zero modes in the internal disk is then 
\beq
\#_{zero \ modes \ disk} = 2 N  R_{bag \, 2} \frac{ g B }{2}  = 2 N n_2 \ .
\eeq
The sum of the disk and the ring gives indeed the correct answer,  $N n$.

Yet another check is provided by studying the  $U(N)$ theory with the number of fundamental scalars $N_{f}$ larger than $ N$.  The scalar potential is of the form, 
\begin{eqnarray}
V =  \frac{g^{2}}{4}  \Tr_{N}\left( q   q^\dagger - \xi {\bf 1}_{N \times N}\right)^2,
 \end{eqnarray}
as a natural extension of (\ref{nonabelianvortex}),  
 where $ q $ now is an $N \times N_f$ matrix.  Inside the vortex bag, the scalar fields take the form, 
 \beq  q = \left(\begin{array}{cccccc}0 &  &  &  & 0 & \hdots \\ & \sqrt\xi  &  &  & \vdots &  \\ &  & \ddots &  & \vdots &  \\ &  &  & \sqrt{\xi} & 0 & \hdots\end{array}\right)
 \eeq
Expansion of the potential $V$ around such values of $q$ determines
the masses of the scalar fields inside the vortex. It is obvious that
the negative mass squared terms can only arise from the part 
\beq   \frac{g^{2}}{4}  (  \sum_{A=1}^{N_{f}} q_{1}^{A}  {\bar   q_{1}^{A} } - \xi )^{2} =   -   \frac{g^{2} \xi}{2}    \sum_{A}^{N_f}     q_{1}^{A}  {\bar  q_{1}^{A} } + \ldots
\label{negative} \eeq
in  the  $(11)$ element of $\left( q   q^\dagger - \xi {\bf 1}_{N\times N}\right)^2$,  as all other terms contain positive coefficients. 
However, the terms $A=2,\ldots, N$ in (\ref{negative})  are exactly canceled by terms arising from the product of nondiagonal elements 
\begin{eqnarray}    &&  \frac{g^{2}}{4}  \sum_{A, B} \sum_{j=2}^{N} [  q_{1}^{A}  {\bar  q_{j}^{A} } q_{j}^{B}  {\bar q_{1}^{B}  }    + (1 \leftrightarrow j)   ]  
\to     \frac{g^{2}}{4}  \sum_{A=2}^{N} [  q_{1}^{A}  (\sqrt{\xi } + {\bar  q_{A}^{A} })  (\sqrt{\xi} + q_{A}^{A}  ) {\bar  q_{1}^{A} } +\ldots    \nonumber  \\
&=&     \frac{g^{2} \xi }{2}  \sum_{A=2}^{N}   q_{1}^{A}   {\bar  q_{A}^{1}} +\ldots\;. 
\end{eqnarray}
so that the tachyonic scalars, with  mass squared  $ -\tfrac{g^{2} \xi }{2}$ are  $q_{1}^{1}$ and $ q_{1}^{A}$, $A= N+1, \ldots, N_{f}$.    
According to the discussion of the beginning of this section, taking
into account the magnetic field inside the bag, $\tfrac{g\xi}{2}$,
(we consider the BPS case, $\lambda=1$)  the number of the scalar zero
modes is then $1 +  N_{f}-N$.   Adding the vector zero modes which are
unchanged: ($N-1$), one finds a total of $N_{f}$, or by taking into
account the Landau level degeneracy: $N_{f}\, n$ zero modes.

BPS non-Abelian vortices for  $N_{f}> N$ are `semilocal'  vortices:  the modulus  contains the vortex transverse size moduli, and their structure is 
very rich and interesting  \cite{Hanany:2003hp,Shifman:2006kd,Eto:2007yv} (see for instance \cite{Eto:2007yv} for a new, Seiberg-like duality in pairs of systems of different $(N_{f}, N)$'s having closely related moduli spaces).    In any event, the dimension of the moduli space can be deduced very generally e.g.,  from an index theorem or from  the symplectic quotient construction of the moduli space \cite{Hanany:2003hp,Eto:2007yv}:  
\beq \{{\bf Z}, {\bf \Psi}, \tilde{\bf \Psi} | D=0 \} /U(n)  \;; \label{symplecticquotient}
\qquad  D=[{\bf Z}^\dagger,{\bf Z}]+ {\bf \Psi}^\dagger {\bf \Psi} -{\bf \tilde{\Psi}}{\bf \tilde{\Psi}}^{\dagger}- \xi\;, 
\eeq
where  ${\bf Z}$, ${\bf \Psi}$ and  ${\bf \tilde{\Psi}}$ are $n\times n$, $N \times n$, and $n \times (N_{f}-N)$  matrices, respectively.  Its (complex) dimension is therefore given by
\beq   n^{2}+  nN +  n (N_{f}-N )  -  n^{2}  = n  N_{f},  
\eeq
in agreement with the zero mode counting.

Our last example of nontrivial checks refers to the cases with different coupling constants for the Abelian and non-Abelian  gauge group factors.
From now on we focus on the case with $N=2$. The $U(2)$ gauge field can be decomposed as 
\beq
A_{\mu} =  \frac{a_{\mu}}{2} {\bf 1} + \frac{A^a_{\mu}}{2} \sigma^a \ ,
\eeq
and the covariant derivative is 
\beq
D_{\mu} = \partial_{\mu} - i \frac{e a_{\mu}}{2} {\bf 1}   - i   \frac{g A_{\mu}^a }{2}\sigma^a \ .
\eeq
The choice of different couplings is very natural, especially if one
considers the fact that $g$ has a quantum mechanical running distinct
from that of the Abelian one, $e$, and can be tuned to be equal to the latter only at a specific energy scale.  This case was considered in the very first paper  \cite{Auzzi:2003fs}.

The BPS Lagrangian for arbitrary $e$ and $g$  is 
\bea   \label{arbitrary} 
{\cal L} &=& -\frac{1}{4}f_{\mu\nu}f^{\mu\nu}-\frac{1}{4}F_{\mu\nu}^a
F^{\mu\nu a} + \Tr \, (D_\mu q)^{\dagger}(D^\mu q) \nonumber \\
&& -   \frac{e^2}{8} \left( |q|^2 - 2 {\xi}\right)^2-   \frac{g^2}{8}
\sum_a \Tr \left(q^\dag \sigma^a q\right)^2 \ ,
\eea
where $|q|^2={\rm Tr}\, (q q^{\dagger})$. The BPS equations are
\bea
f_{xy} +  \frac{e}{2} \left( |q|^2- 2 {\xi}\right)&=& 0   \;; \nonumber \\
F_{\mu\nu}^a + \frac{g}{2} \Tr \, q^\dag \sigma^a q &=& 0   \;;  \nonumber \\
(D_x + i D_y ) q &=& 0 \;.
\eea
We derive things in a different order than we did before.
First we search the stable vacuum which would then correspond to the interior phase of the multi-vortex. 
A solution of the  BPS equations is the following magnetic phase
\beq
\label{magneticphase}
q = \left(\begin{array}{cc}
0&\\
&e \sqrt{\frac{2 \xi}{e^2 + g^2}}\\
\end{array}\right) \ ,  \qquad \quad f_{xy} = \frac{ e g^2\xi}{e^2 + g^2} \ , \qquad \quad F^3_{xy} = \frac{ e^2 g \xi}{e^2 + g^2}  \ .
\eeq
This is the internal phase of the non-Abelian vortex for generic couplings.
For $e=g$ this reduces to (\ref{qinsidena}) and (\ref{binsidena}).

 \begin{figure}[b]
\centerline{
\epsfxsize=7.0cm \epsfbox{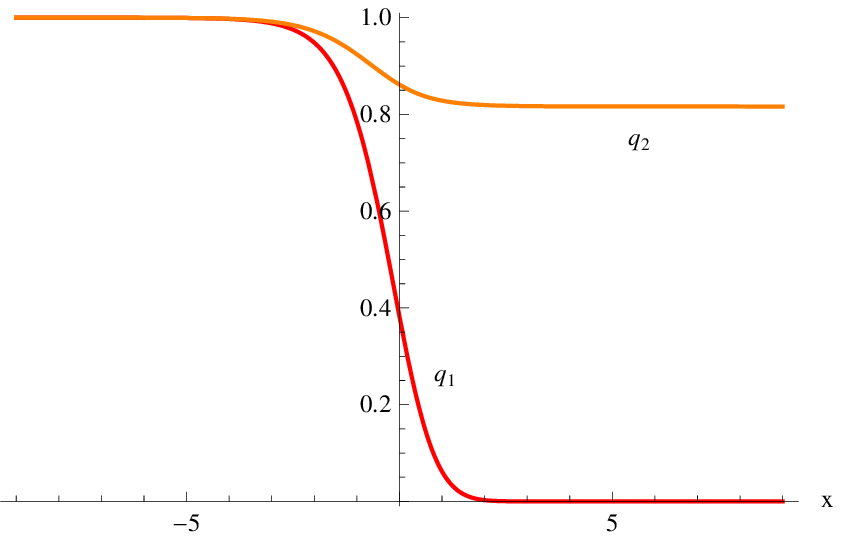}\qquad \quad
\epsfxsize=7.0cm \epsfbox{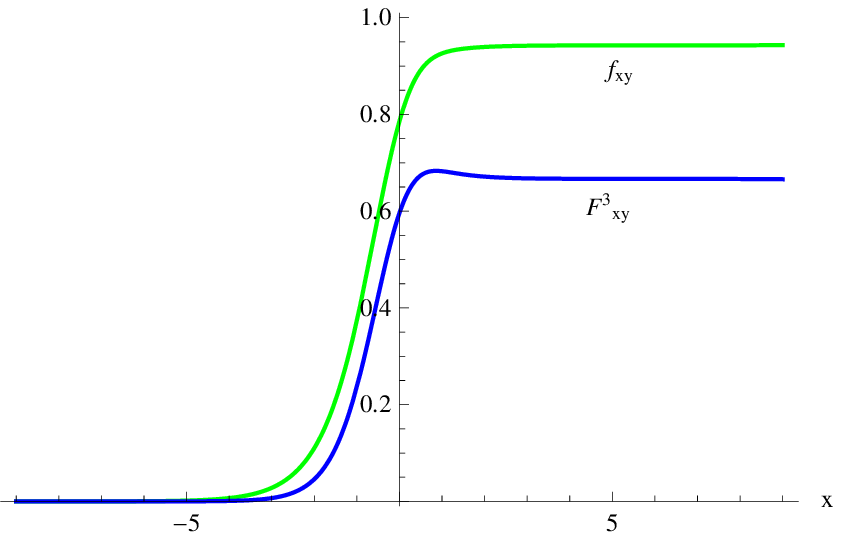}}
\caption{{\footnotesize Domain wall solution of the equations (\ref{bpseqdiffcouplins}) for different couplings. The figure refers to the values $e=1$ and $\gamma =2$.}}
\label{walldiffcouplings}
\end{figure}

We construct the domain wall between the Higgs phase and the magnetic
phase using the following Ansatz 
\beq 
a_{y}(x)\ , \qquad   A_y^3(x)\ , \qquad 
q = \left(\begin{array}{cc}
q_1(x)&\\
&q_2(x)\\
\end{array}\right) \ ,
\eeq
and the BPS equations become
\bea
{a_y}' +  \frac{e}{2} \left(q_1^2 + q_2^2  - 2\xi\right) &=& 0 \ ,\nonumber \\
{A^3_y}' +  \frac{g}{2}  \left( q_1^2 - q_2^2 \right) &=& 0 \ , \nonumber \\
{q_1}' +  \left(\frac{e}{2} a_y + \frac{g}{2}   A_y^3 \right) q_1 &=& 0 \ , \nonumber \\
{q_2}' + \left(\frac{e}{2} a_y - \frac{g}{2}  A_y^3 \right) q_2 &=& 0 \ .
\eea
These then reduce to the following two coupled second-order equations:
\bea
\label{bpseqdiffcouplins}
(\log q_1)'' &=& \frac{e^2}{4}\left( (1+\gamma) q_1^2 + (1 - \gamma) q_2^2  -2  \xi \right)  \;; \nonumber \\
(\log q_2)'' &=&  \frac{e^2}{4} \left( (1-\gamma) q_1^2 + (1 + \gamma) q_2^2  -2 \xi\right) \;,  
\eea
where $\gamma = g^2/e^2$. The magnetic fields are related to the scalar fields by
\beq
f_{xy} =  \frac{e}{2} \left(2 \xi - q_1^2 -q_2^2\right) \ , \qquad F_{xy}^3 = \frac{g}{2}\left(q_2^2 - q_1^2\right) \ .
\eeq
A   domain wall solution interpolating between the Higgs and magnetic phases is given by the numerical solution in Figure \ref{walldiffcouplings} for the case $\gamma=2$.
The case of equal couplings $\gamma=1$ is simpler  because $ q_2 = 1$ and $
f_{xy} =  F_{xy}^3$. Another simplification occurs in the non-Abelian strong coupling limit $\gamma \to \infty$ for which it can be seen that a solution is given by $q_1 = q_2$ and $ F_{xy}^3 = 0$.

The multi-vortex is an area of the magnetic phase (\ref{magneticphase}) separated from the Higgs phase by the previously found domain wall. 
The $W$ bosons, when expanded around the vacuum of the magnetic phase,
have the mass squared
\beq
m_W^2 = \frac{g^2e^2 \xi}{e^2+g^2} \ ,
\eeq
which is the generalization of (\ref{massWbosoninside}) to unequal $U(1)$ and $SU(N)$ gauge couplings.
Given the non-Abelian magnetic field $F^3_{xy}$  in (\ref{magneticphase}), this is exactly the value for the lowest level to be marginal.

As for the scalars, expansion of the scalar potentials in Eq.~(\ref{arbitrary}) around the value of $q$ in Eq.~(\ref{magneticphase})  gives   the quadratic terms 
\beq      -   \frac{ e^2  g^2  \xi  }{ e^2 + g^2}   q^1_1  (q^1_1)^*   +   \frac{e^2 \xi}{4}    (q^2_2 +  (q^2_2)^* )^2\;.     \label{q11only}
\eeq  
The only tachyonic scalar is  $q^1_1 $.   
Now   by making the replacement
\beq      e B  \to      \frac{e f_{xy} +  g F_{xy}^3 }{2}= \frac{e^2  g^2 }{e^2 + g^2}   \xi \;,         
\eeq
in Eq.~(\ref{scalarenergy}) and Eq.~(\ref{spectrumscalar1}), where
we used the values of the magnetic fields (\ref{magneticphase}),
one gets for the spectrum of $q^1_1$ 
\beq 
E_{n_1, n_2} = \frac{e^2  g^2 \xi  }{e^2 + g^2} (2 \,n_2 +1) + m^2 \;:    
\eeq
we see that the negative mass squared  $  m^2 = -   \frac{ e^2  g^2  \xi  }{ e^2 + g^2} $  in  (\ref{q11only})   is precisely the value  which gives  the zero energy modes.

\section{Discussion  \label{Discussion}}

A close relationship is thus found to exist between the general vortex
zero modes and magnetic instabilities of the type discussed by
Ambj{\o}rn, Nielsen and Olesen. The large flux limit, in which the vortex
interior has an almost constant magnetic field, is an ideal setup for
disclosing such a connection. We used the $W$-boson gyromagnetic
instability and similar ones for the scalar and fermion fields. The
counting of zero modes obtained this way  and the  dimension of
the known moduli spaces of BPS non-Abelian vortices match precisely in 
all cases and provides a unifying picture,  valid for translational,
orientational, fermionic or semilocal zero modes. This seems to be
particularly remarkable in view of the fact that the way the Landau-level
zero-point energy,  the Zeeman term and the  mass term  add up to zero
is different for various  types of fields.  We conclude that there is
a universal mechanism for the generation of the vortex zero modes,
which encompasses both the onset of Ambj{\o}rn, Nielsen, Olesen magnetic
instabilities in the electroweak theory (or in QCD), and all sorts of
vortex zero modes inherent in Abelian and non-Abelian vortices.

Let us clarify that the fact that our counting of the vortex zero modes coincides, in the case of BPS vortices,  with the known dimension of the vortex moduli space as well as  with the known index theorem,  just shows that 
our analysis is correct and consistent. Even though it is quite nontrivial to show how things work out,  leading to such a consistent picture,  this is not the main purpose of our analysis.   

  Most importantly, it was shown for the first time that  the BPS vortex configuration reduces, in the large winding limit,  precisely to the critical situation envisaged by Olesen and Ambj{\o}rn,  which corresponds to the onset of magnetic instabilities of the broken phase of e.g., standard Weinberg-Salam theory.  
  
 This was quite unexpected and surprising,  as  the magnetic instability analyses in \cite{Nielsen:1978rm,Ambjorn:1988tm,Ambjorn:1989bd}  were made  in the partially broken phase of e.g., $SU(2)_{L}\times U_Y(1)$  theory with unbroken 
 $U_{EM}(1)$ gauge group. The authors of  \cite{Nielsen:1978rm,Ambjorn:1988tm,Ambjorn:1989bd} then considered some external magnetic source which produces a strong external magnetic field   of the unbroken   $U_{EM}(1)$.    
 This is quite in contrast to the standard setting of non-Abelian vortices, where one considers the vacuum 
  in a fully Higgsed phase, i.e., with no massless gauge bosons in the bulk.  The orientational zero modes arise in the latter   due to the presence of the global color-flavor diagonal symmetry (absent in systems considered in  \cite{Nielsen:1978rm,Ambjorn:1988tm,Ambjorn:1989bd}), broken by individual vortex solutions. 
  Therefore the two classes of systems  look quite distinct and it would seem hardly possible to find any contact between the two.  
  
 What was shown here is that  actually the two seemingly unrelated
 physics phenomena,  the Nielsen-Olesen-Ambj{\o}rn magnetic
 instabilities  and non-Abelian vortices, are deeply related by the
 universal mechanism of charged zero modes in the presence of magnetic
 fields. To prove such a connection, the consideration of the large winding limit of the latter turned out to be particularly useful. 
 
   Such a close connection found here then brings us to comment on  some physics interpretation emphasized by Ambj{\o}rn and Olesen. 
In a somewhat unrealistic BPS saturated version of electroweak theory,
with $ \lambda= \tfrac{g^2}{8 \sin^2 \theta}$, where $\theta$ is the Weinberg angle and $\lambda$ is the quartic Higgs coupling, these authors find the first order  (BPS) equation  \cite{Ambjorn:1988tm} , 
\beq f_{12} =    \frac{g}{2 \sin \theta} \phi_0^2 + 2 \sin \theta |w|^2, \label{antiscreening}
\eeq
and an analogous equation for $Z_{12}$,  where $f_{12}$  is the
$U_{EM}(1)$ magnetic field, $\phi_0$ is the Higgs VEV. The second term
on the right-hand side is then interpreted as an "antiscreening
effect",   where the condensate of the $W$ bosons  tends to increase
the applied magnetic field $f_{12}$, in contrast to what happens in
the ANO vortex  (where the scalar condensate tends to diminish the
magnetic field -  screening effect, or Lenz's law).   It is then natural to ask  \footnote{We thank Poul Olesen for raising this question to us (private communication).} whether the non-Abelian vortices show screening  or antiscreening effect. 

As a non-Abelian vortex is in a sense  simply  an ANO vortex {\it embedded} in a particular color-flavor corner,  the standard screening effect is certainly there.   As for the "antiscreening effect", Eq.~(\ref{antiscreening}),  a color-flavor rotation (orientational zero modes) is accompanied by the excitation of the $W^{\pm}$ boson components of the vortex configuration, 
see Eq.~(\ref{corners}-\ref{perfect}), in exactly the same mechanism that brings us  to Eq.~(\ref{antiscreening}). 
 Therefore one might conclude that the non-Abelian vortices possess both 
screening (scalar condensates) and anti screening effect ($W$ boson condensates).  

These  considerations,  at the same time, lead us  to an alternative interpretation of the second term of   Eq.~(\ref{antiscreening}).  Namely, the fact that  the $W$ bosons become  massless at the critical magnetic field due to the Zeeman effect means that the $SU_L(2)$  symmetry is  (at least locally around the vortex)  restored.  Now the electromagnetic  gauge field is 
\beq   A_{\mu} =  \sin \theta  \, W^3_{\mu} + \cos \theta \, B_{\mu},  
\eeq
where $W^{(a)}_{\mu}$ and $B_{\mu}$  stand for $SU_L(2)$ and $U_Y(1)$  gauge bosons, respectively.  In the broken $SU_L(2)$ phase, the $U_{EM}(1)$ magnetic field is  then 
\beq    f_{12} =  \sin \theta   ( \partial_1   W^3_{2}  - \partial_2  W^3_{1})   + \cos \theta (\partial_1   B_{2}  - \partial_2  B_{1})\;,   
\eeq 
whereas in the unbroken  phase the  $SU_L(2)$    field tensor is  given by   
 \beq     W_{12}^3 =     \partial_1   W^3_{2}  - \partial_2  W^3_{1}  +  \epsilon^{3ab} W^{a}_1  W^{b}_2 =   \partial_1   W^3_{2}  - \partial_2  W^3_{1}  - 2\, |w|^2,
\eeq
where  the form of the condensate   Eq.~(\ref{profile}) for 
\beq W^- =\tfrac{1}{\sqrt{2}}   (W^1 - i W^2) \eeq 
 has been used.   At this point it is quite clear that Eq.~(\ref{antiscreening}) simply  signals the fact that  the equation of motion is  being satisfied by  $f_{12}$, in which the Abelian 
 tensor  $ \partial_1   W^3_{2}  - \partial_2  W^3_{1}$  is    replaced by a non-Abelian $SU_L(2)$ tensor,  $  \partial_1   W^3_{2}  - \partial_2  W^3_{1}  +  \epsilon^{3ab} W^{a}_1  W^{b}_2$. 
 The analogous term on the $Z_{12}$ equation can also be understood as the restoration of non-Abelian nature of $SU_L(2)$  fields.    
 
Such a reinterpretation  is very much  in line with the result of Ambj{\o}rn and Olesen \cite{Ambjorn:1989bd}  that the  magnetic instability  and vortex formation   at the critical $U_{EM}(1)$   magnetic field is actually nothing but the onset of  phase transition to the unbroken  $SU_L(2)\times U_Y(1)$  symmetric phase  of the electroweak theory.

\section*{Acknowledgments} The authors thank Poul Olesen for raising interesting questions about the 
non-Abelian vortices and the Nielsen-Olesen-Ambj{\o}rn magnetic instabilities,  which triggered the present investigation.
We thank Jarah Evslin for useful discussions.
The  work of S.B.  is funded by the Grant ``Rientro dei Cervelli Rita Levi Montalcini'' of the Italian government.   
 S.B.  wishes to thank E. Rabinovici for discussions on magnetic instabilities.  S.B.G.   thanks Institute of Modern Physics, Lanzhou, China,  for
hospitality. The present research is supported by the INFN special project GAST 
(``Gauge and String Theories").

\end{document}